\documentclass[12pt]{article}
\usepackage{amssymb}
\def\beq{\begin{equation}}
\def\eeq{\end{equation}}
\usepackage[all,2cell,dvips]{xy}

\def\IR{\relax{\rm I\kern -.18em R}}

\begin{document}
\title{On the non-abelian superalgebra spanned by the conserved quantities of $N=1$ supersymmetric Korteweg-de Vries equation }
\author{ \Large S. Andrea*, A. Restuccia**, A. Sotomayor***}
\maketitle{\centerline {*Departamento de Matem\'{a}ticas,}}

\maketitle{\centerline{Universidad Sim\'on Bol\'{\i}var}}
\maketitle{\centerline{**Departamento de F\'{\i}sica}}
\maketitle{\centerline{Universidad de Antofagasta}
\maketitle{\centerline{**Departamento de F\'{\i}sica}}
\maketitle{\centerline{Universidad Sim\'on Bol\'{\i}var }

\maketitle{\centerline{***Departamento de Matem\'aticas}
\maketitle{\centerline{Universidad de Antofagasta}}
\maketitle{\centerline{e-mail: sandrea@usb.ve, arestu@usb.ve,
sotomayo81@yahoo.es }}
\begin{abstract}We obtain an infinite sequence of bosonic non-local conserved quantities for the $N=1$ supersymmetric  Korteweg-de Vries
equation. It is generated from a bosonic non-local conserved
quantity of Super Gardner equation. In distinction to the already
known one with odd parity and dimension $\frac{1}{2}$, it has
even parity and dimension $1$. It fits exactly in the
supersymmetric cohomology in the space of conserved quantities
that we also introduce here. Using results from this cohomology
we obtain the Poisson bracket of several non-local conserved
quantities, including the already known odd ones and the new even
ones. The algebra closes in terms of polynomials of local and
non-local conserved quantities. We prove that the bosonic
non-local conserved quantities cannot be expressed as functions
of the already known local and non-local conserved quantities of
Super KdV equation.
\end{abstract}


Pacs: 11.30.Pb, 12.60.Jv, 02.30.Ik, 11.30.-j

\section{Introduction}Non-local conserved quantities are relevant in
order to analyze algebraic and non-perturbative aspects of field
theory \cite{Luscher,Brezin,Vega}. From the algebraic point of
view the interesting non-local conserved quantities should
classically span an algebra in terms of Poisson brackets. If this
algebraic structure extends to the quantum formulation then it
will determine important properties of the field theory.

It is known \cite{Bazhanov} that the conformal algebra, generated
by the energy momentum tensor $T$ with the composite fields built
as powers of $T$ and its derivatives, contains an infinite
dimensional abelian subalgebra whose classical limit is spanned
by the local conserved quantities of KdV equation, see also
\cite{Sasaki,Eguchi,Kupershmidt}.

It is then of interest to look for an algebraic extension of this
abelian subalgebra, which should then be formulated in terms of
non-local conserved quantities.

It is known that $N=1$ Super KdV \cite{Mathieu,Manin} possesses an
infinite set of odd non-local conserved quantities \cite{Dargis}.
They may be derived from the Super Gardner equation \cite{Mathieu}
following the original idea of Gardner \cite{Gardner}. There are
two known conserved quantities of Super Gardner equation. The
local one \cite{Mathieu} and an odd non-local one found in
\cite{Andrea3}. They generate all known local and non-local
conserved quantities of Super KdV equation. A Gardner deformation
has also been used to obtain the $N=2,a=4$ KdV hierarchy from
Kaup-Boussinesq equation \cite{Kiselev}.

 In this paper we present a bosonic non-local conserved quantity
of Super Gardner equation, it generates an infinite sequence of
bosonic non-local conserved quantities of Super KdV equation.
These ones cannot be expressed as functions of the previously
known local and non-local conserved quantities. We then analyze
the algebra spanned by all local and non-local conserved
quantities of Super KdV. We do so by introducing an algebraic and
analytic approach which we call the supersymmetric cohomology.
From it we can obtain several of the Poisson brackets involving
the non-local conserved quantities. The bracket of two non-local
conserved quantities contains non-linear terms, polynomials in
lower dimensional conserved quantities. This non-abelian algebra
of non-local conserved quantities is an unexpected feature of KdV
theory. It involves a consistent construction of Poisson brackets
of non-local quantities, an area of recent interest in
mathematical-physics.

\section{The $N=1$ Super KdV equation and the Gardner map} We denote $\Phi$ a superfield,
$\Phi:\mathbb{R}\rightarrow\Lambda$ where $\Lambda$ is a
Grassmann algebra with one generator singled out,
\[\Phi=\xi(x)+\theta u(x),\] where $u(x)$ and $\xi(x)$ have even
and odd parity respectively.

We consider the ring of Schwartz superfields
\[C^\infty_\downarrow(\mathbb{R},\Lambda)=\left\{\Phi \in
C^\infty( \mathbb{R},\Lambda)/\lim_{x\rightarrow \pm
\infty}x^p\frac{\partial^q}{\partial x^q}\Phi=0\right\}, \] for
every $p,q\geq 0$, the ring of integrable superfields
\[C^\infty_I(\mathbb{R},\Lambda)=\left\{\Phi \in
C^\infty(\mathbb{R},\Lambda)/\frac{\partial}{\partial
\theta}\Phi\in C^\infty_\downarrow(\mathbb{R},\Lambda)\right\},\]
and the rings of non-local superfields
\[C^\infty_{NL,1}(\mathbb{R},\Lambda)=\left\{\Phi \in C^\infty(
\mathbb{R},\Lambda)/D\Phi\in
C^\infty_\downarrow(\mathbb{R},\Lambda) \right\},\]
\[C_{NL,2}^\infty(\mathbb{R},\Lambda)=\left\{\Phi \in
C^\infty( \mathbb{R},\Lambda)/D^2\Phi\in
C^\infty_\downarrow(\mathbb{R},\Lambda) \right\},\] where
$D=\frac{\partial}{\partial\theta}+\theta\frac{\partial}{\partial
x }$.

On the integration formulas in the next section we explicitly use
that $C_\downarrow^\infty$ is an ideal of $C_{NL,2}^\infty$ and
$C^\infty_{NL,1}$.

The following relation holds $C^\infty_\downarrow\subset
C^\infty_{NL,1} \subset C^\infty_I.$

In order to introduce non-local quantities we denote
\[D^{-1}\left(A+\theta B\right)=\int_{-\infty}^xB(s)ds+\theta
A(x)\] for any superfield $A+\theta B \in C_\downarrow^\infty$ .
The map $D^{-1}:C_\downarrow^\infty\rightarrow C_{NL,1}^\infty$
satisfies $DD^{-1}=D^{-1}D=I$.

The Super KdV equation \cite{Mathieu} in superfield form is \beq
\Phi_t=D^6\Phi+3D^2(\Phi D\Phi),\label{susyeq}\eeq and the
corresponding Gardner \cite{Gardner} equation in terms of the
superfield $\chi$ is \beq\chi_t=
 D^6\chi+3D^2(\chi D\chi)-3\epsilon^2
 (D\chi)D^2(\chi D\chi).\label{superg}\eeq
 (\ref{susyeq}) and (\ref{superg}) are related by the Super
 Gardner map \cite{Mathieu}
 \beq \Phi=\chi+\epsilon D^2\chi-\epsilon^2\chi D
 \chi.\label{strag}\eeq

  It gives \beq
\begin{array}{l}\Phi_t-D^6\Phi-3D^2\left(\Phi D\Phi
\right)=\left[1+\epsilon D^2-\epsilon^2\left(D\chi+\chi
D \right) \right]\cdot \\
\left\{\chi_t-D^6\chi-3D^2\left(\chi D\chi
\right)+3\epsilon^2\left(D\chi \right)D^2\left(\chi D\chi \right)
\right\}.\end{array}\label{factsupercampos}\eeq (\ref{strag}) maps
solutions of (\ref{superg}) into solutions of (\ref{susyeq}).

The inverse is also true, provided we include in the space of
solutions of (\ref{superg}) those which are formal series in
$\epsilon,$ $\chi=\sum_{n=0}^\infty
a_n(\Phi,D\Phi,...)\epsilon^n.$ Under the assumption $\chi\in
C_\downarrow^\infty$, (\ref{strag}) implies $\Phi\in
C_\downarrow^\infty.$

\section{A bosonic non-local conserved quantity of Super Gardner equation}
We introduce in this section a bosonic non-local conserved
quantity of the Super Gardner equation. We denote it $H_G$, \beq
H_G=\frac{1}{2}\int dxd\theta
D^{-1}\left\{D\left[\frac{\exp\left(\epsilon D^{-1}\chi
\right)-1}{\epsilon}\right] D^{-1}\left[\frac{\exp\left(-\epsilon
D^{-1}\chi\right)-1}{\epsilon}\right]\right\},\label{consgard2}\eeq
assuming that $\chi\in C_\downarrow^\infty.$

  $H_G$ is a well
defined integral. In fact, the factor
$D\left[\frac{\exp\left(\epsilon D^{-1}\chi
\right)-1}{\epsilon}\right]\in C_\downarrow^\infty $ while the
factor $D^{-1}\left[\frac{\exp\left(-\epsilon
D^{-1}\chi\right)-1}{\epsilon}\right]\in C_{NL,2}^\infty, $ hence
the product belongs to $C_\downarrow^\infty$. Finally
\[D^{-1}C_\downarrow^\infty\subset C_{NL,1}^\infty\subset
C_I^\infty,\]and consequently the integral is a well defined
quantity.

$H_G$ has the same parity as the local conserved quantity and
opposite to the already known non-local conserved quantity of
Super Gardner equation.

We now prove that $H_G$ is a conserved quantity of the Super
Gardner equation.

We denote $f(\epsilon)\equiv \frac{1}{\epsilon}\left(\exp
(\epsilon D^{-1}\chi)-1\right)$, then
$f(-\epsilon)=\frac{1}{\epsilon}\left(1-\exp(-\epsilon
D^{-1}\chi)\right)$.

Then
\[\frac{dH_G}{dt}=-\frac{1}{2}\int dxd\theta D^{-1}\left[D\partial_tf(\epsilon)\cdot D^{-1}f(-\epsilon)+
Df(\epsilon)\cdot D^{-1}\partial_tf(-\epsilon)\right].\] The dot
$\cdot$ in the above and following expressions is used to denote
the product of superfields. It is omitted when the expressions do
not lead to ambiguities.

 The evaluation of $\partial_tf(\epsilon)$ on the
solutions of the Super Gardner equation yields
\[\partial_t f(\epsilon)=\exp(\epsilon D^{-1}\chi)\cdot \left(D^{-1}\frac{\partial \chi}{\partial t}\right)=D\left(g(\epsilon)\right)\]
where \[g(\epsilon)=\exp \left(\epsilon
D^{-1}\chi\right)\cdot\left(F_0+\epsilon
F_1+\epsilon^2F_2\right)\] and $F_0=D^4\chi+3\chi D\chi,F_1=\chi
D^3\chi-D\chi D^2\chi,F_2=-2\chi{\left(D\chi\right)}^2.$

Thus \[\frac{dH_G}{dt}=-\frac{1}{2}\int dxd\theta
D^{-1}\left[D^2g(\epsilon)\cdot
D^{-1}f(-\epsilon)+Df(\epsilon)\cdot g(-\epsilon) \right],\]
moreover the integrand may be rewritten as
\[D^{-1}\left[D^2\left(g\left(\epsilon\right)D^{-1}f(-\epsilon)\right)-g(\epsilon)Df(-\epsilon)+Df(\epsilon)\cdot g(-\epsilon)\right].\]
Consequently,
\[\frac{dH_G}{dt}=-\frac{1}{2}\int dxd\theta D\left[g(\epsilon)D^{-1}f(-\epsilon)+2\chi D^2\chi\right]\]
with the bracket term belonging to $C_\downarrow^\infty(
\mathbb{R},\Lambda).$ In fact, $g(\epsilon)\in
C_\downarrow^\infty( \mathbb{R},\Lambda)$ and
$D^{-1}f(-\epsilon)\in C_{NL,2}^\infty ( \mathbb{R},\Lambda).$

We then conclude that $\frac{dH_G}{dt}=0$ on the solutions of the
Super Gardner equation: $H_G$ is a conserved quantity of the
Super Gardner equation.

$H_G$ may be rewritten as \beq \begin{array}{ll} H_G=\int
dxd\theta\left\{D^{-1}\left[\frac{\exp\left(\epsilon
D^{-1}\chi\right)+\exp\left(-\epsilon D^{-1}\chi
\right)-2}{2\epsilon^2} \right]\right.+  \\ \left.
+\frac{1}{2}\left[\frac{\exp\left(\epsilon D^{-1}\chi
\right)-1}{\epsilon}\right] D^{-1}\left[\frac{\exp\left(-\epsilon
D^{-1}\chi\right)-1}{\epsilon} \right]
\right\}.\end{array}\label{consgard1}\eeq Although each term in
the integrand does not belong to $C_I^\infty,$ the summation of
both terms is in $C_I^\infty$.

Although the presentation of the conserved quantity in this
section has been in a top down scheme, it was really obtained the
other way around. We got the first two bosonic non-local conserved
quantities of SKdV equation from a recursive approach and later
one, having them as reference, we deduced using the Gardner
transform the conserved quantity of the Super Gardner equation.
In the following section on the supersimmetric cohomology we
explain why this bosonic quantities should arise.

\section{Infinite sequence of bosonic non-local conserved quantities of Super KdV equation}
We now follow the ideas in \cite{Gardner} and in \cite{Mathieu}.

Following \cite{Gardner} the conserved quantity $H_G$ induces
then infinitely many non-local conserved quantities of
(\ref{susyeq}). The first two of them are:

\beq H^{NL}_1=-\frac{1}{2}\int dxd\theta D^{-1}(\Phi D^{-2}\Phi),
\label{ncq1} \eeq

\beq \begin{array} {ll}H_3^{NL}=\int dxd\theta
D^{-1}\left[\Phi\left(-\frac{1}{2}D^2\Phi+
 D\Phi\cdot D^{-2}\Phi+\right.\right.\\\left.\left.+\frac{1}{2}D^{-2}\Phi\cdot
{\left(D^{-1}\Phi\right)}^2+\frac{1}{4}D^{-1}\Phi\cdot
D^{-1}{\left(D^{-1}\Phi\right)}^2 \right)\right].\end{array}
\label{ncqsteve}\eeq

(\ref{ncq1}) and (\ref{ncqsteve}) may be rewritten as

\beq H_1^{NL}=\int dxd\theta
\left[D^{-1}\left(\frac{1}{2}{\left(D^{-1}\Phi\right)}^2\right)-\frac{1}{2}D^{-1}\Phi
\cdot D^{-1}\left(D^{-1}\Phi\right)\right],\label{cohom7}\eeq

\beq\begin{array}{ll} H^{NL}_3=\int
dxd\theta\left[-\frac{1}{2}D^{-1}(\Phi D^2\Phi)+D^{-1}(D\Phi\cdot
\Phi \cdot D^{-2}\Phi)+\frac{1}{24}D^{-1}{(D^{-1}\Phi)}^4-\right.
\\\left.
-\frac{1}{6}D^{-2}\Phi{(D^{-1}\Phi)}^3+\frac{1}{8}{(D^{-1}\Phi)}^2D^{-1}{(D^{-1}\Phi)}^2\right].\end{array}
\label{ncq2}\eeq

Each term in the integrands of (\ref{cohom7}) and (\ref{ncq2})
belongs to $C_{NL,2}^\infty$, but the combination is in
$C_I^\infty$.

It is interesting to rewrite them in terms of component fields,
for example, (\ref{ncq1}) becomes
\beq-\frac{1}{2}\int_{-\infty}^\infty
dx\left(\xi(x)\int_{-\infty}^xds\xi(s)\right).\label{incomp}\eeq
It is straightforward to verify that the time derivative of
(\ref{incomp}) evaluated on the solutions of SKdV is zero.

The infinite sequence of local conserved quantities of the $N=1$
Super KdV equation were obtained by Mathieu \cite{Mathieu} from
one conserved quantity of the Super Gardner equation \[G_1=\int
dxd\theta\chi=\sum_{n=0}\epsilon^{2n}H_{2n+1},\] using the
inverse Gardner transformation.
 The infinite sequence of fermionic non-local
conserved quantities were obtained by Dargis and Mathieu
\cite{Dargis} and Kersten \cite{Kersten1}. Dargis and Mathieu used
a recursion operator method. This infinite sequence was latter on
obtained from one non-local conserved quantity of the Super
Gardner equation
\[G_{\frac{1}{2}}^{NL}=\int dxd\theta \left(\frac{\exp(\epsilon
D^{-1}\chi)-1}{\epsilon}\right)=\sum_{n=0}\epsilon^nH_{n+\frac{1}{2}}^{NL}\]
in \cite{Andrea3}.

In this paper we introduce the bosonic non-local conserved
quantity in (\ref{consgard2}) which may be denoted $G_1^{NL}$. The
subindex denotes the dimension of the corresponding conserved
quantity. It is interesting that both non-local conserved
quantities $H_{2n+\frac{3}{2}}^{NL}$ and $G_1^{NL}$ arise from the
fermionic and bosonic parts respectively of the superfield
\[\frac{1}{2}\int dx D\left[\frac{\exp\left(\epsilon D^{-1}\chi
\right)-1}{\epsilon}\right] D^{-1}\left[\frac{\exp\left(-\epsilon
D^{-1}\chi\right)-1}{\epsilon}\right]\] (notice that there is no
$\theta$ integration in the formula). In fact
\begin{eqnarray*}&&\frac{1}{2}\int dx D\left[\frac{\exp\left(\epsilon
D^{-1}\chi \right)-1}{\epsilon}\right]
D^{-1}\left[\frac{\exp\left(-\epsilon
D^{-1}\chi\right)-1}{\epsilon}\right]=G_1^{NL}+\theta\left[\sum_{n=0}H_{2n+\frac{3}{2}}^{NL}\epsilon^{2n}\right]+
\\&&+\frac{\theta}{2}\left[\frac{\exp \epsilon
G_1-1}{\epsilon}\right]\cdot
\left[\sum_{n=0}{(-1)}^{n+1}\epsilon^n
H_{n+\frac{1}{2}}^{NL}\right]\end{eqnarray*} (we thank an unknown
referee for his remark on this point).

By taking \[\frac{1}{2}\int dx d\theta
D\left[\frac{\exp\left(\epsilon D^{-1}\chi
\right)-1}{\epsilon}\right] D^{-1}\left[\frac{\exp\left(-\epsilon
D^{-1}\chi\right)-1}{\epsilon}\right]\] one only projects the
$\theta$ factor in the above expression while by taking
\[\frac{1}{2}\int dx d\theta D^{-1} \left[D\left[\frac{\exp\left(\epsilon D^{-1}\chi
\right)-1}{\epsilon}\right] D^{-1}\left[\frac{\exp\left(-\epsilon
D^{-1}\chi\right)-1}{\epsilon}\right]\right]\] one projects the
$\theta$ independent part.

$G_{\frac{1}{2}}^{NL}$ and $G_1^{NL}$ are both non-local
conserved quantities of Super Gardner equation and are not
related by any algebraic relation. In fact, if we take the lowest
dimensional conserved quantity arising from $G_1^{NL}$ we obtain
(\ref{incomp}). It is not trivial and it cannot be constructed as
a function of the Dargis and Mathieu conserved quantities. At the
end of section 6 we give a general proof of this statement.

The property that the $2n+\frac{3}{2}$ sector of
$G_{\frac{1}{2}}^{NL}$ and $G_1^{NL}$ arises from an associated
superfield, is similar to the case of the lowest dimensional
conserved quantity obtained by Dargis and Mathieu $\int dxd\theta
D^{-1}\Phi$ and the lowest dimensional local quantity $\int
dxd\theta \Phi.$ Both are constructed from the fermionic and
bosonic projections respectively of the superfield $\int dx \Phi
$, where only integration in $x$ is considered. Moreover, the
$2n+\frac{1}{2}$ sector of $G_{\frac{1}{2}}^{NL}$ and the local
conserved quantities $H_{2n+1}$ arise from an associated
superfield in an analogous way.

\section{The SUSY cohomology on the space of conserved quantities}
The invariance under supersymmetry of SKdV equations implies that
the SUSY transformations of conserved quantities are also
conserved quantities. That is, if $H=\int dxd\theta h,h\in
C_I^\infty,$ is conserved under the SKdV flow then
\[\delta_QH:=\int dxd\theta Qh\] is also a conserved quantity,
where $Q=-\frac{\partial}{\partial \theta}+\theta
\frac{\partial}{\partial x}$ is the generator of supersymmetric
transformations and anticommutes with the covariant derivative
$D$.

The operation $\delta_Q$ acting on functionals of the above form
is well defined since the equivalence class of integrands
\[h\rightarrow h+Dg,\] with $g\in C_\downarrow^\infty$, leaving
$H$ invariant, is transformed under $\delta_Q$ to the equivalence
class of $Qh$
\[Qh\rightarrow Qh+QDg=Qh+D(-Qg)\] where $Qg\in
C_\downarrow^\infty$.

$\delta_Q$ is a superderivation satisfying $\delta_Q\delta_Q=0.$
In fact, \[\delta_Q\delta_QH=\int dx d\theta Q^2h=-\partial_\theta
h|_{-\infty}^\infty=0\] since $h\in C_I^\infty.$

For the local conserved quantities of SKdV, which we denote
$H_{2n+1}(\Phi),n=0,1,\ldots,$ we have
\beq\delta_QH_{2n+1}(\Phi)=0,n=0,1,\ldots,\label{cohom0}\eeq
where the index $2n+1$ denotes the dimension of $H_{2n+1}.$

If we consider the ring $C_\downarrow^\infty$ of superfields,
$H_{2n+1}$ is closed but not exact. However if we extend the ring
to the superfields in $C_{NL,1}^\infty$,
$C_\downarrow^\infty\subset C_{NL,1}^\infty,$ then $H_{2n+1}$
becomes exact and it is expressed in terms of
$\delta_QH_{2n+\frac{1}{2}},n=0,1,\ldots$ where
$H_{n+\frac{1}{2}},n=0,1,\ldots$ denote the odd non-local
conserved quantities of SKdV \cite{Dargis,Kersten1}, they have
dimension $n+\frac{1}{2}$. The remaining
$H_{2n+\frac{3}{2}}^{NL},n=0,1,\ldots$ plus a polynomial of lower
dimensional conserved quantities is closed but not exact in
$C_{NL,1}^\infty$, however if we extend the ring of superfields
to $C_{NL,2}^\infty$ they become exact and equal to
$\delta_QH_{2n+1}^{NL}$, where $H_{2n+1}^{NL}$ are the even
non-local conserved quantities we have introduced in the previous
section. They have dimension $2n+1$. To obtain the exact relation
between them we use the conserved quantities of the Super Gardner
equation.

We denote them $G_1,G_{\frac{1}{2}}^{NL}$ and $G_1^{NL}.$ We have
\beq G_1=\int
dxd\theta\chi=\sum_{n=0}\epsilon^{2n}H_{2n+1},\label{cohom1}\eeq
\beq G_{\frac{1}{2}}^{NL}=\int dxd\theta \left(\frac{\exp(\epsilon
D^{-1}\chi)-1}{\epsilon}\right)=\sum_{n=0}\epsilon^nH_{n+\frac{1}{2}}^{NL},
\label{cohom2}\eeq \beq  \begin{array}{ll} H_G\equiv
G_1^{NL}=\int dxd\theta\left\{D^{-1}\left[\frac{\exp\left(\epsilon
D^{-1}\chi\right)+\exp\left(-\epsilon D^{-1}\chi
\right)-2}{2\epsilon^2} \right]\right.+  \\ \left.
+\frac{1}{2}\left[\frac{\exp\left(\epsilon D^{-1}\chi
\right)-1}{\epsilon}\right] D^{-1}\left[\frac{\exp\left(-\epsilon
D^{-1}\chi\right)-1}{\epsilon} \right]
\right\}=\sum_{n=0}\epsilon^{n}H_{n+1}^{NL},\end{array}
\label{cohom3}\eeq where $\chi \in C_\downarrow^\infty$.

We notice that the odd powers of $\epsilon$, in(\ref{cohom3}),
are polynomials expressions of the Dargis and Mathieu conserved
quantities of Super KdV. For example \beq
H_2^{NL}=\frac{1}{2}H_{\frac{1}{2}}^{NL}H_{\frac{3}{2}}^{NL}.
\label{secquanttriv}\eeq

We start the analysis by considering \[\delta_QG_1=\int dxd\theta
Q\chi=\chi|_{-\infty}^{+\infty}=0\] since $\chi\in
C_\downarrow^\infty$, hence we obtain (\ref{cohom0}).

We also have \beq
\begin{array}{cc}\delta_QG_{\frac{1}{2}}^{NL}=\frac{\exp(\epsilon G_1)-1}{\epsilon}=G_1+\frac{1}{2}\epsilon G_1^2+\cdots
\\ =\sum_{n=0}\epsilon^{2n}H_{2n+1}+\frac{1}{2}\epsilon{\left(\sum_{n=0}\epsilon^{2n}H_{2n+1}\right)}^2+\cdots\end{array}\label{cohom6} \eeq
from which we obtain the relation between the odd non-local and
local conserved quantities. In particular we get \beq
\delta_QH_\frac{1}{2}^{NL}=H_1, \label{cohom4}\eeq and
\[\delta_QH_\frac{3}{2}^{NL}=\frac{1}{2}H_1^2=\delta_Q(\frac{1}{2}H_1H_\frac{1}{2}^{NL})\]
that is
\beq\delta_Q\left(H_\frac{3}{2}^{NL}-\frac{1}{2}H_1H_\frac{1}{2}^{NL}
\right)=0.\label{cohom5}\eeq

This is the generic situation , from (\ref{cohom6}),
$H_{2n+1},n=0,1,\ldots$ is expressed as an exact quantity in
terms of $\delta_Q$[$H_{2n+\frac{1}{2}}^{NL}+\Sigma$ products of
lower dimensional conserved quantities] while
[$H_{2n+\frac{3}{2}}^{NL}+\Sigma$ products of lower dimensional
conserved quantities] is closed in the ring $C_{NL,1}^\infty.$ If
we extend the ring of superfields to $C_{NL,2}^\infty$, then the
closed quantity becomes exact and expressed in terms of
$H_{2n+1}^{NL},n=0,1,\ldots $ The integrand of $H_{2n+1}^{NL}$ is
expressed in terms of superfields in $C_{NL,2}^\infty$ with the
property that the whole integrand belongs to $C_I^\infty.$ In the
case of $H_{n+\frac{1}{2}}^{NL}$ the integrand is expressed in
terms of superfields in $C_{NL,1}^\infty\subset C_I^\infty,$
hence each term is integrable. For example let us consider
$H_1^{NL}$ (see (\ref{cohom7})). The integrand is in terms of
$D^{-1}h$ where $h$ are the integrands of previously known
conserved quantities $H_{n+\frac{1}{2}}^{NL}$ and $H_{2n+1}$. In
this particular case we have
\[H_{\frac{3}{2}}^{NL}=\int \frac{1}{2}dxd\theta{(D^{-1}\Phi)}^2,\]
\[H_1=\int dxd\theta \Phi,\] \[H_{\frac{1}{2}}^{NL}=\int dxd\theta D^{-1}\Phi.\]
This is also a generic property of $H_{2n+1}^{NL},n=0,1,\ldots$
and as it was already known of
$H_{n+\frac{1}{2}}^{NL},n=0,1,\ldots$ whose integrands may be
expressed in terms of polynomials in $D^{-1}h$ where $h$ are the
integrands of the local conserved quantities $H_{2n+1}$.

From (\ref{cohom7}) we have
\[\delta_QH_1^{NL}=H_{\frac{3}{2}}^{NL}-\frac{1}{2}H_1H_{\frac{1}{2}}^{NL}\]
that is, the closed quantity becomes exact in $C_{NL,2}^\infty$.
Similar relations are obtained from (\ref{cohom3}) for higher
dimensional conserved quantities. The general formula is
\beq\sum_{n=0}\epsilon^{n}\delta_QH_{n+1}^{NL}=\sum_{n=0}\epsilon^{2n}H_{2n+\frac{3}{2}}^{NL}-\frac{1}{2\epsilon}
\left[\exp\left(\sum_{n=0}\epsilon^{2n+1}H_{2n+1}\right)-1\right]
\left[\sum_{n=0}{(-\epsilon)}^{n}H_{n+\frac{1}{2}}^{NL}\right]\label{genform}\eeq
In order to obtain (\ref{genform}) we used the non trivial
property
\begin{eqnarray*}& &\exp^{\epsilon D^{-1}\chi}=\Sigma_0^\infty \epsilon^np_n\\
 & & \exp^{-\epsilon D^{-1}\chi}=\Sigma_0^\infty {(-\epsilon)}^np_n+D\Sigma_0^\infty\epsilon^n\Psi_n, \end{eqnarray*}
 for certain odd superfields $\Psi_n\in C_\downarrow^\infty(
 \mathbb{R},\Lambda).$ The integrals of $p_n$ are obtained from
 (\ref{cohom2}).

We then have the following relations between the conserved
quantities of SKdV equation:

 \beq\xymatrix {H_1  &  &  H_3 &  & H_5  &  &H_7 &\cdots &
 \\ H_{\frac{1}{2}}^{NL}\ar[u]
   &H_{\frac{3}{2}}^{NL}  &  H_{\frac{5}{2}}^{NL}\ar[u] &H_{\frac{7}{2}}^{NL} & H_{\frac{9}{2}}^{NL}\ar[u] &\ldots
 &  &  &
 \\ &  H_1^{NL}\ar[u] &  & H_3^{NL}\ar[u] &  & H_5^{NL}\ar[u] & & } \label{diagram}\eeq
 where the arrow denotes the action of $\delta_Q$, up to lower
 dimensional conserved quantities as explained previously.

 The bosonic conserved quantities $H_1^{NL},H_3^{NL},\ldots$ fit then
 exactly in the SUSY cohomology of the previously known conserved
 quantities.
 \section{The Poisson bracket of the conserved quantities}
 We consider the Poisson bracket of superfields $\Phi$ at two
 different points in superspace as given in \cite{Mathieu}:
 \beq
\left\{\Phi(x_1,\theta_1),\Phi(x_2,\theta_2)\right\}=P_1\Delta,\label{brac1}
\eeq  where
$P_1=D_1^5+3\Phi_1D_1^2+(D_1\Phi_1)D_1+2(D_1^2\Phi_1)$ and
$\Delta=\delta(x_1-x_2)(\theta_1-\theta_2).$

We now use this fundamental bracket and the Super Leibnitz rules
satisfied by the Poisson bracket to obtain the Poisson bracket of
local and non-local conserved quantities of SKdV equation. The
first evaluation of Poisson brackets of non-local conserved
quantities of Super KdV equation was performed in \cite{Dargis}.
We complete their work including the non linear terms which
appear in the evaluation of the brackets.

At each step in the evaluation we have to verify that the
integrands belongs to $C_I^\infty$.

It is well known that the local conserved quantities $H_{2n+1}$
commute with any other conserved quantity. We obtain for the
Poisson bracket(commutators or anticommutators) according to
parity between $H_{\frac{1}{2}}^{NL}$ and any non-local conserved
quantity, the following: let

\[I\equiv\left\{\int d x_1d\theta_1D^{-1}\Phi_1,\int dx_2d\theta_2\left(h\left(\Phi_2,D_2\Phi_2,\ldots,D_2^{-1}\Phi_2\right)\right)\right\}.\]
 We have \[\int dxd\theta D^{-1}\Phi=\int dxd\theta
\left(\theta\Phi\right).\]

Using the superderivative property of the Poisson bracket we
obtain
\begin{eqnarray*} I&=&\int dx_1d\theta_1\int dx_2d\theta_2
\theta_1\left\{\Phi_1,h \right\}=
\\& =& \int dx_1d\theta_1
\int dx_2d\theta_2 \theta_1\left[\left\{\Phi_1,\Phi_2
\right\}\frac{\partial h}{\partial \Phi_2}+\left\{\Phi_1,D_2\Phi_2
\right\}\frac{\partial h}{\partial
D_2\Phi_2}\right.+\\&&\left.+\cdots+\left\{\Phi_1,D_2^{-1}\Phi_2
\right\}\frac{\partial h}{\partial D_2^{-1}\Phi_2}\right]= \\
& =& \int dx_1d\theta_1\int
dx_2d\theta_2\theta_1\left[\left(P_1\Delta\right)\frac{\partial h
}{\partial \Phi_2}-D_2\left(P_1\Delta\right)\frac{\partial h
}{\partial D_2\Phi_2}+ \cdots-\right.\\ &-& \left.
D_2^{-1}\left(P_1\Delta\right)\frac{\partial h }{\partial
D_2^{-1}\Phi_2}\right]= \\ & =&\int dx_1d\theta_1 \int
dx_2d\theta_2 P_1\theta_1\left[\Delta\cdot\frac{\partial
h}{\partial \Phi_2 }+D_2\Delta\cdot\frac{\partial h}{\partial
D_2\Phi_2}+\cdots D_2^{-1}\Delta\cdot\frac{\partial h}{\partial
D_2^{-1}\Phi_2} \right],
\end{eqnarray*}
where $P_1\theta_1=Q_1\Phi_1$ and
\[D_2\Delta=-D_1\Delta,D_2^2\Delta=-D_1^2\Delta,D_2^3\Delta=D_1^3\Delta,D_2^4\Delta=D_1^4\Delta,D_1^{-1}\Delta=D_2^{-1}\Delta+1.\]
Replacing these relations into the above integral we get
\begin{eqnarray*} I&=&\int dx_1d\theta_1\int dx_2d\theta_2\left[\left(Q_1\Phi_1\right)\Delta
\frac{\partial h}{\partial
\Phi_2}+\left(D_1Q_1\Phi_1\right)\Delta\frac{\partial h }{\partial
D_2\Phi_2}+ \right.
\\ &+&\left.
\left(D_1^2Q_1\Phi_1\right)\Delta\frac{\partial h }{\partial
D_2^2\Phi_2}+\cdots
D_1^{-1}\left(Q_1\Phi_1\right)\Delta\frac{\partial h }{\partial
D_2^{-1}\Phi_2 }\right]=
\\& =&\int dx_1d\theta_1\int
dx_2d\theta_2\Delta\left[Q_1\Phi_1\frac{\partial h}{\partial
\Phi_2 }+\left(Q_1D_1\Phi_1\right)\frac{\partial h}{\partial
D_2\Phi_2}+\right.\\ &+& \left.
\left(Q_1D_1^2\Phi_1\right)\frac{\partial h}{\partial D_2^2\Phi_2
}+\cdots+
 \left(Q_1D_1^{-1}\Phi_1\right)\frac{\partial
h }{\partial D_2^{-1}\Phi_2}\right]=
\\ & =&-\int dxd\theta
Qh\left(\Phi,D\Phi,\ldots,D^{-1}\Phi\right)=-\delta_QH.
\end{eqnarray*}

We remark that each step in the calculation of the integrals is
well defined. Since we are working with a distribution $\Delta$,
we assume in this calculation that $\Phi$ is of compact support.

The above argument, valid for the $H_{n+\frac{1}{2}}^{NL}$
conserved quantities may be extended to any conserved quantity
$H$. In fact, \beq
\left\{\Phi_1,H(\Phi_2)\right\}=P_1\lim_{\epsilon\rightarrow 0
}\frac{H(\Phi_2+\epsilon\Delta)-H(\Phi_2)}{\epsilon}\eeq for any
$H(\Phi).$ We then have
\begin{eqnarray*}\left\{H_{\frac{1}{2}}^{NL},H\right\}&=&\int_1dx_1d\theta_1\left\{D_1^{-1}\Phi_1,H\right\}=\\
&=&\int dx_1d\theta_1\theta_1\left\{\Phi_1,H\right\}=\int
dx_1d\theta_1\theta_1P_1\lim_{\epsilon\rightarrow 0
}\frac{H(\Phi_2+\epsilon\Delta)-H(\Phi_2)}{\epsilon}=\\ &=&-\int
dx_1d\theta_1P_1\theta_1\lim_{\epsilon\rightarrow 0
}\frac{H(\Phi_2+\epsilon\Delta)-H(\Phi_2)}{\epsilon}=\\ &=&-\int
dx_1d\theta_1 Q_1\Phi_1\lim_{\epsilon\rightarrow 0
}\frac{H(\Phi_2+\epsilon\Delta)-H(\Phi_2)}{\epsilon}=-\delta_QH.
\end{eqnarray*}

We thus obtain
\begin{eqnarray}& & \left\{H_{\frac{1}{2}}^{NL},H_{2n+1}\right\}=-\delta_QH_{2n+1}=0,   \\
& &
\left\{H_{\frac{1}{2}}^{NL},H_{n+\frac{1}{2}}^{NL}\right\}=-\delta_QH_{n+\frac{1}{2}}^{NL},
\\ & &
\left\{H_{\frac{1}{2}}^{NL},H_{n+1}^{NL}\right\}=-\delta_QH_{n+1}^{NL}.
\end{eqnarray}
For example,
\begin{eqnarray}& & \left\{H_{\frac{1}{2}}^{NL},H_{\frac{1}{2}}^{NL}\right\}=H_1   \\
& &
\left\{H_{\frac{1}{2}}^{NL},H_{\frac{3}{2}}^{NL}\right\}=\frac{1}{2}H_1^2
\\ & &
\left\{H_{\frac{1}{2}}^{NL},H_1^{NL}\right\}=-\frac{1}{2}H_1H_{\frac{1}{2}}^{NL}+H_{\frac{3}{2}}^{NL}
\end{eqnarray}
where $\delta_QH_{n+\frac{1}{2}}$ and $\delta_QH_{n+1}$ are
obtained from (\ref{cohom6}) and (\ref{genform}).

We now calculate the Poisson bracket
$\left\{H_{\frac{3}{2}}^{NL},H_{\frac{3}{2}}^{NL}\right\}. $ We
have
\[\left\{H_{\frac{3}{2}}^{NL},H_{\frac{3}{2}}^{NL}\right\}=\int_1dx_1d\theta_1\int_2dx_2d\theta_2
\left\{D_1^{-1}\Phi_1,D_2^{-1}\Phi_2\right\}D_1^{-1}\Phi_1D_2^{-1}\Phi_2\]
where
\[\left\{D_1^{-1}\Phi_1,D_2^{-1}\Phi_2\right\}=-D_1^{-1}D_2^{-1}\left\{\Phi_1,\Phi_2\right\}=D_1^{-1}P_1D_2^{-1}\Delta.\]
We obtain
\[D_1^{-1}\Delta=H(x_1-x_2-\theta_1\theta_2)\] where $H(z)$ is the
Heaviside step function: $H(z)=1,z>0$ and $H(z)=0$ if $z<0$ and
explicitly
$H(x_1-x_2-\theta_1\theta_2)=H(x_1-x_2)-\delta(x_1-x_2)\theta_1\theta_2.$

Also
\[D_2^{-1}\Delta=H(x_1-x_2-\theta_1\theta_2)-1,\] and after some
calculations
\begin{eqnarray*}
D_1^{-1}P_1D_2^{-1}\Delta&=&D_1^2\delta(x_1-x_2-\theta_1\theta_2)+D_1\left[2\Phi_1H(x_1-x_2-\theta_1\theta_2)-2\Phi_1\right]+
\\ &&
+D_2\left[2\Phi_2H(x_1-x_2-\theta_1\theta_2)\right]+\Phi_1\Delta=\left(D_1^3+\Phi_1\right)\Delta+
\\ &&
+D_1\left[2\Phi_1H(x_1-x_2-\theta_1\theta_2)-2\Phi_1\right]+D_2\left[2\Phi_2H(x_1-x_2-\theta_1\theta_2)\right].
\end{eqnarray*}

Using these expresions in the above Poisson bracket we get
\begin{eqnarray*}\left\{H_{\frac{3}{2}}^{NL},H_{\frac{3}{2}}^{NL}\right\} &=&
\int_1dx_1d\theta_1\int_2dx_2d\theta_2\left(D_1^3+\Phi_1\right)\Delta
D_1^{-1}\Phi_1D_2^{-1}\Phi_2= \\ &=& \int dxd\theta\left(D^2\Phi
D^{-1}\Phi+\Phi D^{-1}\Phi D^{-1}\Phi \right)= \\ &=&\int
dxd\theta \left(-\Phi
D\Phi+D\left(\frac{1}{3}{\left(D^{-1}\Phi\right)}^3\right)\right)=-H_3+\frac{1}{3}{(H_1)}^3
\end{eqnarray*}

 where $H_3$ and $H_1$ are the local conserved
quantities defined in the previous section.

We notice that besides the expected conserved quantities arising
from dimensional considerations, there are terms which are
polynomials in the lower dimensional conserved quantities.

We have thus obtain several relations pointing to the existence of
an algebra, under the Poisson bracket, of conserved quantities
which extends the trivial algebra of local conserved quantities.
This algebra is expected to play a relevant role in the
quantization of the Super KdV hamiltonian.

Related nonlinear structures were observed in \cite{Plyushchay}.

$\smallskip$

We will now use the Poisson algebra to prove that the conserved
quantities $H_{2n+1}^{NL}$ cannot be expressed as a function of
the Dargis and Mathieu and local conserved quantities.

Let us consider a bosonic function
$F(H_{n+\frac{1}{2}}^{NL},H_{2n+1})$ and the bracket
\[\{F(H_{n+\frac{1}{2}}^{NL},H_{2n+1}),H_{\frac{1}{2}}^{NL}\}.\]
We have, using the results before equation (24),
\begin{eqnarray*}&&\left\{H_{2n+1},H_{\frac{1}{2}}^{NL}\right\}=0,\\&&\left\{H_{2n+\frac{1}{2}}^{NL},H_{\frac{1}{2}}^{NL}\right\}=H_{2n+1}+
P_{2n+1}\left(H_{2m+1}\right),\\&&\left\{H_{2n+\frac{3}{2}}^{NL},H_{\frac{1}{2}}^{NL}\right\}=P_{2n+2}\left(H_{2m+1}\right),
\end{eqnarray*}
where $P_k$ is a polynomial of degree $k\geq 2$ of the arguments
and dimension $k$.

We then have
\[\{F(H_{n+\frac{1}{2}}^{NL},H_{2n+1}),H_{\frac{1}{2}}^{NL}\}=\sum_{n}\frac{\partial F}{\partial H_{n+\frac{1}{2}}^{NL}}
\left\{H_{n+\frac{1}{2}}^{NL},H_{\frac{1}{2}}^{NL}\right\}\]
where $\frac{\partial F}{\partial H_{n+\frac{1}{2}}^{NL}} $ is an
fermionic quantity. The term of lowest degree that one can obtain
from such expression is of degree $\geq 2$. On the other side
using the results in section 6 and equation (20)
\[\left\{H_{2n+1}^{NL},H_{\frac{1}{2}}^{NL}\right\}=H_{2n+\frac{3}{2}}^{NL}+P_{2n+\frac{3}{2}}\left(H_{n+1}^{NL},H_{2n+1}\right)\]
we have a linear term $H_{2n+\frac{3}{2}}^{NL}$ for each $n$.
Consequently $H_{2n+1}^{NL}$ cannot be expressed as
$F(H_{n+\frac{1}{2}}^{NL},H_{2n+1})$. In distinction as we
explain in equation (16) $H_{2n+2}^{NL}$ may indeed be expressed
as a polynomial of the Dargis and Mathieu non-local conserved
quantities.

\section{Conclusions}We found a bosonic non-local conserved quantity of Super
Gardner equation. It has even parity and dimension $1$, in
distinction to the known odd one with dimension $\frac{1}{2}$. It
generates an infinite sequence of bosonic non-local conserved
quantities for $N=1$ Super KdV equation. We then introduced the
supersymmetric cohomology. We proved that the bosonic non-local
conserved quantities $H_{2n+1}^{Nl}$ cannot be expressed as
functions of the already known local and fermionic non-local
conserved quantities of Super KdV equation. The SUSY generator
defines a nilpotent operator on the space of all conserved
quantities into itself. On the ring of $C_\downarrow^\infty$
superfields the local conserved quantities are closed but not
exact. However on the ring of $C_{NL,1}^\infty$ superfields, an
extension of the $C_\downarrow^\infty$ ring, they become exact
and equal to the SUSY transformed of the subset of odd non-local
conserved quantities of the appropriate weight. The remaining odd
non-local ones generate closed geometrical objects which become
exact when the ring is extended to the $C_{NL,2}^\infty$
superfields, and equal to the SUSY transformation of the bosonic
non-local conserved quantities we have obtained. These ones fit
exactly in the SUSY cohomology of the already known conserved
quantities. We finally used these properties to evaluate the
Poisson brackets between non-local conserved quantities. The
bracket contains, besides the expected terms from dimensional
arguments, nonlinear terms, polynomials in lower dimensional
local and non-local conserved quantities.
\newpage \textbf{Acknowledgments}

We would like to thank Professor Mikhail Plyushchay for useful
comments and references.

A. R. would like to acknowledge kind hospitality and financial
support from Physics Dept., University of Antofagasta, Chile.


\begin{thebibliography}{}
\bibitem{Luscher}{M. L\"{u}scher and K. Pohlmeyer, Nucl. Phys.
B 137, 46-54 (1978).}
\bibitem{Brezin}{E. Br\'ezin, C. Itzykson, J. Zinn-Justin and J. B. Zuber,
Phys.Lett.  82B, 442-444 (1979).}
\bibitem{Vega}{H. de Vega, H. Eichenherr and J. M. Maillet, Commun. Math. Phys.
92, 507-524(1984).}
\bibitem{Bazhanov}{V. V. Bazhanov, S. L. Lukyanov and A. B. Zamolodchikov , Commun. Math. Phys.
Volume 177, Number 2, 381-398 (1996).}
\bibitem{Sasaki}{R. Sasaki and I. Yamanaka ,  Adv. Stud. in Pure Math. 16, 271-296(1988).}
\bibitem{Eguchi}{T. Eguchi and S. K. Yang , Phys. Lett. B 224, 372-378(1989).}
\bibitem{Kupershmidt}{B. A. Kupershmidt and P. Mathieu, Phys. Lett. B 227,
245-250 (1989).}
\bibitem{Mathieu}{P. Mathieu, J. Math. Phys. 29, 2499
(1988).}
\bibitem{Manin}{Yu. I. Manin and A. O. Radul, Commun. Math. Phys.
98, 65 (1985).}
\bibitem{Dargis}{P. Dargis and P. Mathieu, Phys. Lett. A 176,
67-74 (1993).}
\bibitem{Gardner}{R. M. Miura, C. S. Gardner, and M. D. Kruskal, J. Math. Phys. 9,
1204 (1968).}
\bibitem{Andrea3}{S. Andrea, A. Restuccia and A. Sotomayor, J. Math.
Phys. 46, 103517 (2005).}
\bibitem{Kiselev}{V. Hussin, A. V. Kiselev, A. O. Krutov and T. Wolf, J. Math.
Phys. 51, 083507 (2010).}
\bibitem{Kersten1}{P. H. M. Kersten,  Phys. Lett. A 134, 25 (1988).}
\bibitem{Plyushchay}{C. Leiva and M. S. Plyushchay, JHEP 0310:069 (2003).}


\end{thebibliography}
\end{document}